\documentclass[12pt,preprint]{aastex}

\shorttitle{A simple model of radiative emission in M87}

\shortauthors{E. Vitanza, V. Teresi, D. Molteni, G. Gerardi}

\begin{document}

\title{A simple model of radiative emission in M87}
\author{E. Vitanza, V. Teresi, D. Molteni, G. Gerardi}

\email {vteresi@unipa.it}

\affil{ Dipartimento di Fisica e Tecnologie Relative,\\
 Universit$\grave{a}$ di Palermo, \\ Viale  delle
 Scienze, Palermo, 90128, Italy }



\begin{abstract}
We present a simple physical model of the central source emission
in the M87 galaxy. It is well known that the observed X-ray
luminosity from this galactic nucleus is much lower than the
predicted one, if a standard radiative efficiency is assumed. Up
to now the main model invoked to explain such a luminosity is the
ADAF (Advection-Dominated-Accretion-Flow) model. Our approach
supposes only a simple axis-symmetric adiabatic accretion with a
low angular momentum together with the bremsstrahlung emission
process in the accreting gas. With no other special hypothesis on
the dynamics of the system, this model agrees well enough with the
luminosity value measured by Chandra.
\end{abstract}


\section{Introduction}

M87 is a widely studied galaxy. Many authors have written about
its globular clusters and nucleus, about the hypothesis of the
existence of a supermassive black hole at its centre, and recently
about its jet from the nucleus \citep{Jordan04, Wilson02}. One of
the most investigated problems on the physics of this galaxy is
the problem of its luminosity, particularly the luminosity of its
nucleus. By virtue of Chandra observations an estimate of the
X-ray nuclear luminosity is about $7 \cdot 10^{40}$ $erg
\hspace{0.2cm} s^{-1}$ \citep{DiMatteo03}. Moreover, the data from
Chandra itself allow to obtain temperature, density and radial
speed profiles of the interstellar medium emitting at X-ray
frequencies inside the accretion radius of the central black hole.
From these quantities it is easy to calculate the Bondi accretion
rate, $\dot{M}_{Bondi} = 0.1 \hspace{0.2cm} M_{\odot}
\hspace{0.2cm} yr^{-1}$ \citep{DiMatteo03}. Supposing a canonical
radiative efficiency ($\eta = 0.1$), the estimated luminosity is
$5 \cdot 10^{44}$ $erg \hspace{0.2cm} s^{-1}$, much higher than
the measured value. A common way of solving this problem is based
on the assumption of the ADAF model \citep{Chen97} for the
accreting gas. In this model a significant part of the energy
produced by viscosity is advected towards the central object and
therefore the fractio of gravitational energy transformed into
emitted radiative energy is much lower than in the canonical
Shakura-Sunyaev model \citep{SS73}. Moreover, to solve the energy
excess problem, some authors have formed the hypothesis that
another large part of the produced energy could be converted into
the kinematic
energy of the matter outflowing from the nucleus by the highly energetic jet
\citep{Wilson02, Marshall02}.\\
In our approach, instead, we consider a steady state,
axis-symmetric (with a low angular momentum) and adiabatic
accretion model. Such a model, by virtue of the low value of
angular momentum, is a little refinement of the standard Bondi
flow. The hypothesis of an influence of core rotation on the X-ray
emission in a large, slowly rotating, elliptical galaxy was
already suggested by \citet{Kley95}. The emission process we
suppose is the electron-ion thermal bremsstrahlung. This type of
emission was already considered in the ADAF modeling of M87
\citep{OzDM01}, but not in the basic Bondi model framework. The
nuclear X-ray luminosity calculated from our model is compatible
with the observed value measured by Chandra. Though this model is
very simple, it gives significant results with respect to the
problem of the radiative emission modeling.

\section{The physical model}

We set up a method to fit to an experimental data (the nuclear
luminosity) using a model that is compatible with the known
framework about the source (whose main ingredients are: no
evidence of high rotation, which implies quasi-spherical
accretion, and bremsstrahlung emission process). Moreover, to
perform such a kind of fit we need just one free parameter (the
flow specific angular momentum), whereas the other variables are
bound by the observed values at the accretion radius.\\
Since we will consider only small angular momentum models, we may
neglect the role of viscosity. Indeed, with the low angular
momentum value we will use, the gas will not rotate more than one
orbit from the accretion radius to the black hole. In the
presented model the rotational speed is so small that the gaseous
structure is rather similar to a spherical one only slightly
crushed in the vertical direction by the low angular momentum
value. To obtain temperature, density and radial speed profiles we
consider a set of three equations, in which the symbols used have
the following meanings: $\rho$, $v$ and $a$ are respectively
density, gas radial speed and sound speed, $H_{disk}$ is the
half-thickness of the disk, $\dot{M}$ is the mass accretion rate,
$\lambda$ is the specific angular momentum of the gas, $R_g$ is
the Schwartzschild radius of the black hole, $\gamma = 5/3$ is the
ratio $c_p/c_v$ between the gas specific heats at constant
pressure and volume, $\rho_{\infty}$ and $a_{\infty}$ are the
density and sound speed values taken at a large distance from the
black hole, in our case the values measured by Chandra at the BH
accretion radius.\\


With these definitions, the used equation system is:

mass conservation equation:

\begin{equation}
4\pi r H_{disk} \rho v=\dot{M}=cost
\label{mass}
\end{equation}

radial momentum equation:

\begin{equation}
v\frac{dv}{dr}=-\frac{1}{\rho}\frac{dP}{dr}-\frac{GM}{(r-R_g)^{2}}+\frac{\lambda^{2}}{r^{3}}
\label{radmom}
\end{equation}

polytropic relation between density and sound speed:

\begin{equation}
\rho=\rho_{\infty}\left(\frac{a}{a_{\infty}}\right)^{\frac{2}{\gamma-1}}
\label{polyt}
\end{equation}

Eq. \ref{mass} is the evaluation of the accretion rate $\dot M$
based on the idea that a mass flux $\rho v$ crosses a cylindrical
surface of radius $r$ and height $2H_{disk}$, under the hypothesis
that $\dot M$ is constant in space and time (steady accretion
flow). Eq. \ref{radmom} is the radial momentum transfer equation,
with the lagrangian time derivative of the radial speed (without
the term $\partial v / \partial t$ since we assume a steady state)
in the first member and the three acting forces (pressure
gradient, gravitational and centrifugal) in the second one. Eq.
\ref{polyt} is the thermodynamical relation between density and
sound speed for an ideal gas during an adiabatic process.\\
The disk half-thickness is obtained through the following
procedure. Using the hypothesis of vertical hydrostatical
equilibrium, we can write:

\begin{equation}
\frac{1}{\rho} \frac{\partial P}{\partial z} =
\frac{GM}{(r-R_g)^2} \frac{z}{r}  \label{height0}
\end{equation}

Though the flow we analyze cannot be described as a thin disk, the
only way to evaluate quantitatively the vertical height is to make
this assumption. This means that we approximate the vertical
gradient $\partial P/\partial z$ with $P/H_{disk}$ and substitute
the $z$ value in the right hand side of eq. \ref{height0} with
$H_{disk}$:

\begin{equation}
\frac{1}{\rho} \frac{P}{H_{disk}} = \frac{GM}{(r-R_g)^2}
\frac{H_{disk}}{r} \label{height1}
\end{equation}


By some simple algebraic calculations, this leads to:

\begin{equation}
H_{disk}=(r-R_g)\sqrt{\frac{P}{\rho}\frac{r}{GM}} \label{height}
\end{equation}

We used the Paczynsky-Wiita potential $V_{PW}=-(GM)/(r-R_g)$
\citep{PW} to mimic the general-relativistic gravitational
effects. Note that $v
> 0$ for inflowing gas. The pressure $P$ is
given by $P=\rho a^2/\gamma$.\\
This scheme, containing one differential and two algebraic
equations, can be substituted by a totally algebraic system of
equations by using the Bernoulli relation instead of the radial
momentum differential equation \ref{radmom}:

\begin{equation}
\frac{v^{2}}{2}+\frac{a^{2}}{\gamma-1}-\frac{GM}{r-R_g}+\frac{\lambda^{2}}{2r^{2}}=B
\label{energy2}
\end{equation}

where $B$ is the Bernoulli constant of the gas flow.\\
\\
The algebraic equation system was solved using the following
procedure. By introducing the Mach number $m = v/a$, solving for
$a$ the equation \ref{energy2} and putting all the terms into the
relation \ref{mass}, we obtained an equation in the unknown $m$:

\begin{equation}
\dot{M}=-r\cdot m\cdot a \cdot K \cdot a ^{\frac{2}{\gamma-1}}
  \propto f(m)\cdot A\left( r,B,\lambda \right)
\end{equation}

where $K$ is a constant depending on the entropy of the system and
$f$ is function only of the Mach number:

\begin{equation}
f\left( m\right) =-\frac m{\left[ \frac{m^2}2+\frac
1{\gamma-1}\right]^{ \frac{\gamma +1}{2\left( \gamma -1\right) }}}
\end{equation}

and $A$ is function only of $r$ (B and $\lambda$ are parameters):

\begin{equation}
A\left( r,B,\lambda \right) = r \left[ B- V\left( r,\lambda\right)
\right]^{ \frac{\gamma +1}{2\left( \gamma -1\right) }}
\end{equation}

with $V\left( r,\lambda \right)$, the effective body force
potential (gravitational plus centrifugal), given by:

\begin{equation}
V\left( r,\lambda \right) =\frac{\lambda
^2}{2r^2}-\frac{GM}{r-R_g}
\end{equation}

We used the values of density, flow radial speed and sound speed
$\rho_{\infty}$, $v_{\infty}$ and $a_{\infty}$ at the BH accretion
radius to calculate the Bernoulli constant $B$, necessary to solve
the algebraic system. The unknowns are $\rho$, $v$ and $a$. By
solving for these quantities, we obtained their radial profiles
$\rho(r)$, $v(r)$ and $a(r)$. Finally, from $a(r)$ we found the
temperature profile $T(r)=(m_H a^2)/(2 \gamma K_B)$, where $m_H$
is the proton mass and $K_B$ is the Boltzmann constant. From the
density $\rho$ and the temperature $T$ at a certain radius $r$ we
calculated the emitted power density at the same $r$ for the
bremsstrahlung emission process. Defining $e_{ff}$ as the emitted
power density, $n_{e}$ and $n_{i}$ as the electron and ion
densities in the gas, $Z$ as the atomic number of the ions,
$g_{B}$ as the Gaunt factor, $e$ and $m_{e}$ as the electron
charge and mass, $h$ as the Planck constant, the formula we used
for bremsstrahlung is the following:

\begin{equation}
e_{ff}=(\frac{2\pi K_B}{3m_{e}})^{\frac{1}{2}}\frac{2^5 \pi
e^6}{3hm_{e}c^3}T^{\frac{1}{2}}n_{e}n_{i}Z^{2}g_{B}
\end{equation}

The quantities $n_{e}$ and $n_{i}$ can be calculated from the
density $\rho$ by assuming that the accreting gas is an hot plasma
of fully ionized hydrogen. We adopted in our model the
bremsstrahlung emission mechanism because, as already pointed out,
the dominant emission process for the M87 X-ray nuclear luminosity
is the thermal bremsstrahlung that yields a peak in the X-ray band
\citep{Rey96}. We cut off any emission when the temperature is
larger than $2 \cdot 10^9$ $K$. The main reason for the
temperature cut-off is that, beyond the indicated temperature
limit, the radiation frequency falls in the $\gamma$ band and
therefore the nuclear emission does not contribute to the X-ray
luminosity. It is clear that the procedure we followed is valid
from a physical point of view if the emission process we
considered does not affect very much the flow structure (we did
not include the corresponding terms in the energy equation). This
means that the time-scale of the process should be larger than the
dynamical time of the flow. In the section 3 we show the
comparison among these time-scales.\\
We highlight that, under the
hypothesis of negligible viscosity, the algebraic method we
followed is completely equivalent to the differential equation
approach, since in this case the system is conservative (and the
Bernoulli theorem eq. \ref{energy2} holds). In particular, the
algebraic method allows to find the transonic flow with the sonic
point at the same radius as in the differential equation approach.
In the algebraic scheme the sonic point corresponds to a minimum
of $A(r,B,\lambda)$ as a function of $r$ (for the mathematical
details see the Appendix of \citet{Molteni99}), whereas in the
differential equation approach the sonic point comes out from the
regularity conditions on the function $v(r)$ (i.e. the usual
conditions of numerator and denominator of $dv/dr$ equal to zero).
For conservative systems both methods give the same results
concerning the sonic point position.

\section{Results}

In this section we present the results obtained with a $\lambda$
value, measured in units of $cR_g$ (with $c$ the light speed and
$R_g$ the Schwartzschild radius of the black hole), of 1.555, that
we found to be the value of 'best fit' of the calculated
luminosity to the observed one. This value of angular momentum
gives, at the black hole accretion radius, a rotational speed of
0.93 $km/sec$, that is within the observational error of the
measured speed data \citep{Cohen97}. It is worth to note that,
lowering $\lambda$, the flow structure given by our model gets
closer and closer to the Bondi configuration, until, when
$\lambda$ = 0, it reaches approximately the Bondi model structure.
All the data shown in this section concern the range $3R_g < r < 2
\cdot 10^5 R_g$, since, according to our results, it is within
this region that about the $98 \%$ of the total luminosity is
produced. As we already pointed out in section 2, a criterion to
assess the validity of our model in presence of radiative emission
processes is that the flow is fundamentally adiabatic. This is
verified if the emission time-scale for each considered process is
larger than the flow dynamical time $t_{dyn}=r/v$ at the same
radius. We present in fig. 1 the values of these typical times at
different radii in the range $3R_g < r < 2 \cdot 10^5 R_g$. From
this figure it is clear that the bremsstrahlung emission
time-scale is larger than the flow dynamical time. We do not
consider the synchrotron emission because its frequency range
falls in the radio and (via Comptonization) optical bands
\citep{Rey96}. As regards the soft X-ray emission lines, they have
a significant intensity at low temperatures (about $10^6$ K), that
can be found only beyond the accretion radius.\\
\begin{figure}
\begin{center}
\includegraphics[scale = 0.45,angle = 270.0]{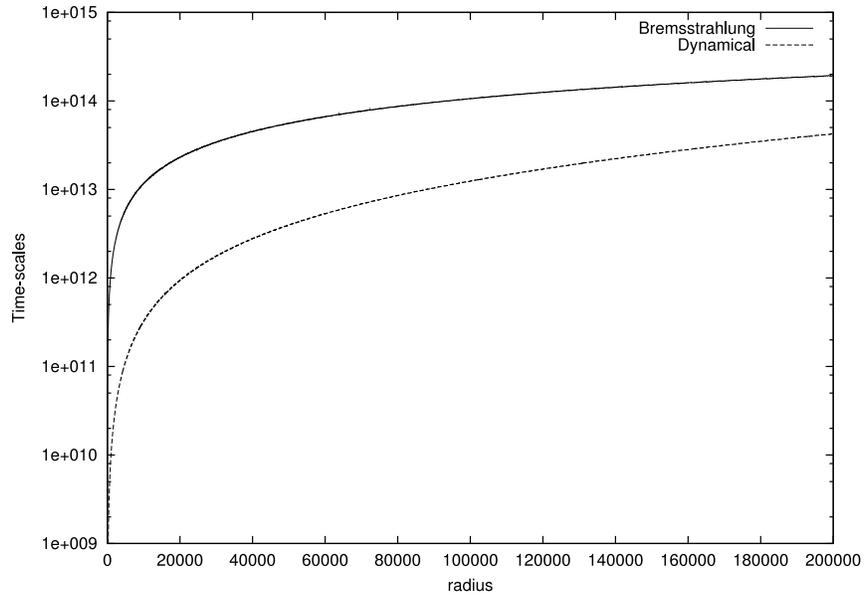}
\caption{Comparison between the bremsstrahlung and the dynamical
time-scales versus $r$. The times are in $secs$.}
\end{center}
\end{figure}
Our model allows to calculate the luminosity emitted by the whole
flow. We present in fig. 2 the partial luminosity $L(r)$ emitted
from $3R_g$ to a generic radius $r$. The figure shows that the
largest part of the total luminosity emitted by the entire flow is
produced in the region from $r = 3R_g$ to $r = 2 \cdot 10^5 R_g$.
The luminosity coming out from the whole system (up to the
external boundary at $r = 5 \times 10^5 R_g$) is $7.1 \cdot
10^{40} \hspace{0.2cm} erg/sec$. Therefore our model permits to
explain the observed luminosity as a result of the accretion flow
emission. Obviously this picture is not the only possible one. For
example, \citet{Wilson02} attribute the origin of the nuclear
X-ray emission to the pc, or sub-pc, scale jet. However our
hypothesis has the advantage of explaining the observational data
in the simple framework of the interstellar medium accreting onto
the central black hole.
\begin{figure}
\begin{center}
\includegraphics[scale = 0.45,angle = 270.0]{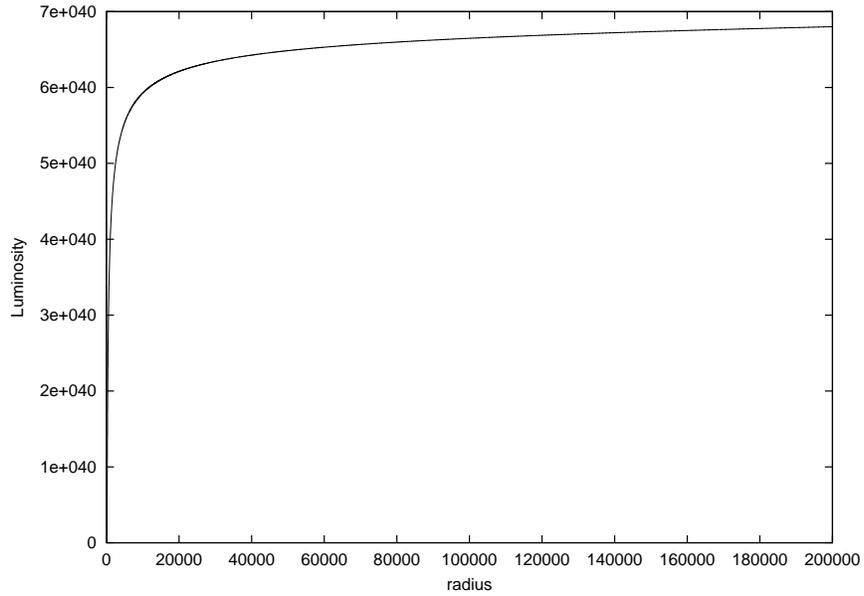}
\caption{Partial luminosity $L(r)$ versus $r$. $L(r)$ is in
$erg/sec$.}
\end{center}
\end{figure}
In figs. 3, 4 and 5 we show the radial profiles of the three
variables that characterize the flow structure: density,
temperature and radial speed. In fig. 6 we show the radial density
of the emitted power for the bremsstrahlung process.
\begin{figure}
\begin{center}
\includegraphics[scale = 0.45,angle = 270.0]{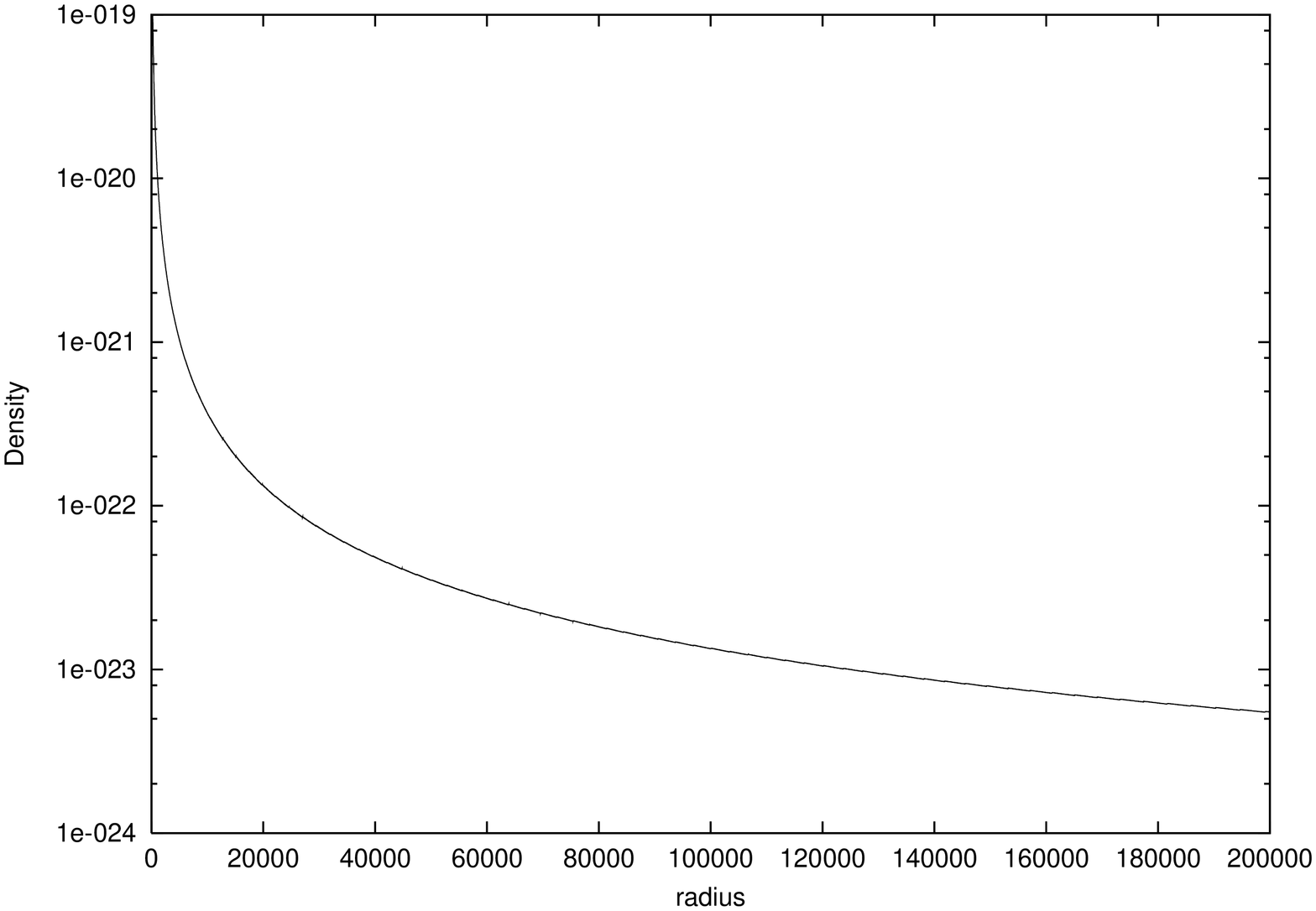}
\caption{Density radial profile. Density is in $g/cm^3$.}
\end{center}
\end{figure}

\begin{figure}
\begin{center}
\includegraphics[scale = 0.45,angle = 270.0]{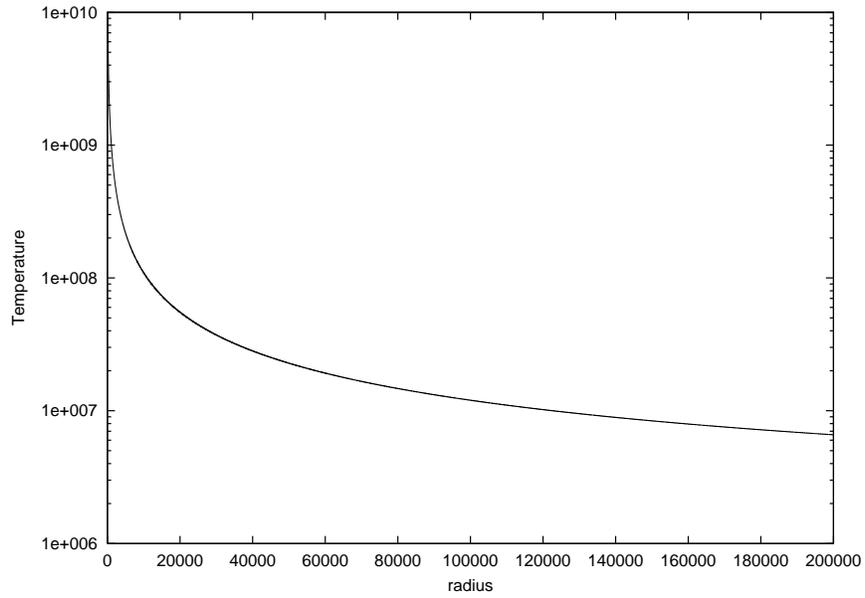} \caption{Temperature radial
profile. The temperature is in $K$ degrees.}
\end{center}
\end{figure}

\begin{figure}
\begin{center}
\includegraphics[scale = 0.45,angle = 270.0]{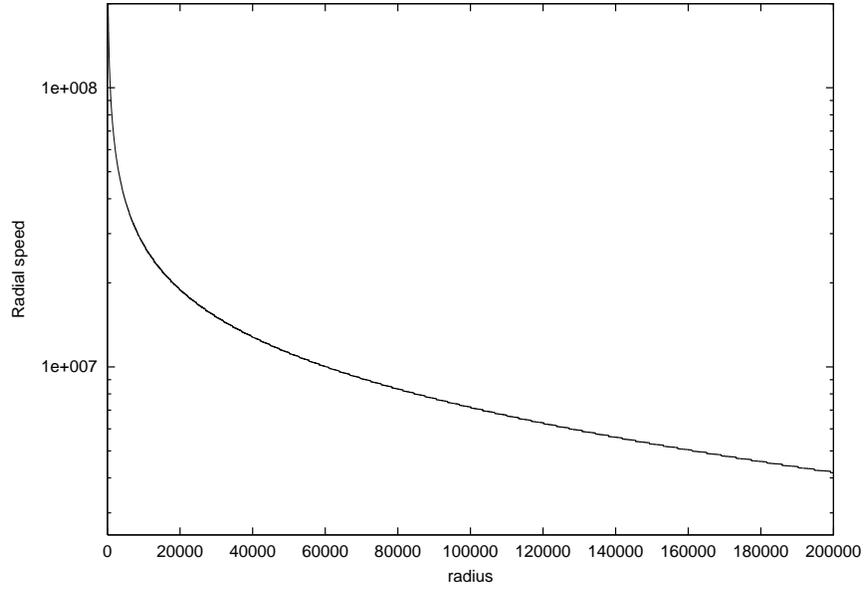}
\caption{Flow radial speed vs. $r$. The speed is in $cm/sec$.}
\end{center}
\end{figure}

\begin{figure}
\begin{center}
\includegraphics[scale = 0.45,angle = 270.0]{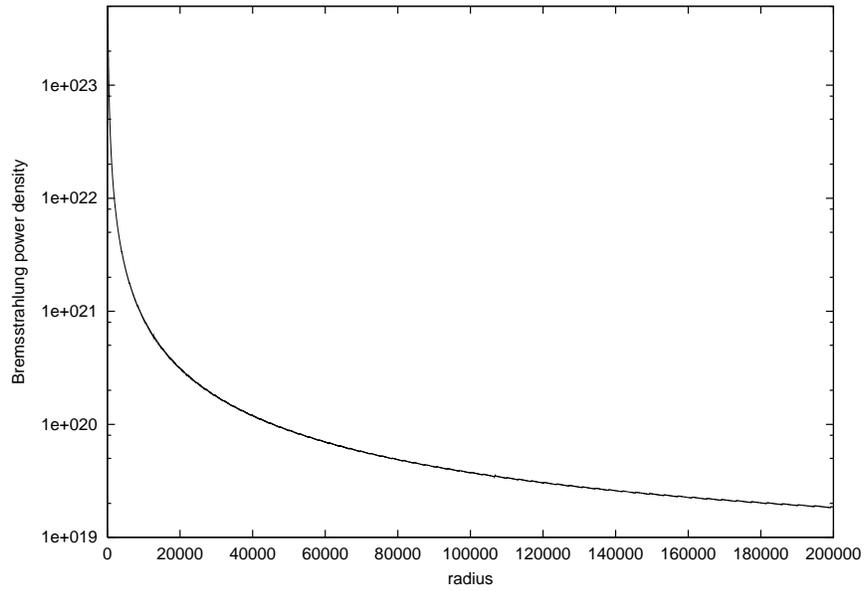}
\caption{Radial density of the emitted power for the
bremsstrahlung process. The power density is in $erg \cdot cm^{-1}
\cdot sec^{-1}$.}
\end{center}
\end{figure}


\section{Conclusions}

In this work we show that the addition of a small gas angular
momentum to a simple adiabatic accretion flow together with the
thermal bremsstrahlung emission process can give, for the active
nucleus of the galaxy M87, a luminosity value that is in good
agreement with the measured one, whereas the value obtained
supposing the standard radiative efficiency is four orders larger
than the measured luminosity. We obtain this result using a very
simple model that contains a new free parameter, the specific
angular momentum $\lambda$ of the accreting gas, that can be
adjusted in order to fit the model to the observed luminosity.
With $\lambda = 1.555$ the obtained luminosity value is $7.1 \cdot
10^{40} \hspace{0.2cm} erg/sec$ versus a measured one of about $7
\cdot 10^{40} \hspace{0.2cm} erg/sec$. Our result can be
considered also a way of giving an estimate of the gas angular
momentum in the nucleus of M87. Moreover, our model could be
applied to other sources in which the low observed luminosity
requires a low radiative efficiency model.

\end{document}